\documentclass[epj]{webofc}
\usepackage[utf8]{inputenc}
\usepackage[varg]{txfonts}   
\usepackage{booktabs}
\usepackage{xcolor}
\definecolor{darkred}{rgb}{0.4,0.0,0.0}
\definecolor{darkgreen}{rgb}{0.0,0.4,0.0}
\definecolor{darkblue}{rgb}{0.0,0.0,0.4}
\usepackage[bookmarks,linktocpage,colorlinks,
    linkcolor = darkred,
    urlcolor  = darkblue,
    citecolor = darkgreen]{hyperref}

\usepackage{amsmath,amssymb,color,amscd,subfigure}
\usepackage{graphicx}
\usepackage{slashed}

\usepackage{subfigure}

\wocname{EPJ Web of Conferences}
\woctitle{Lattice2017}

\newcommand{\ev}[1]{\left < #1 \right >}

\newcommand{\be}{\begin{equation}}
\newcommand{\ee}{\end{equation}}
\newcommand{\bea}{\begin{eqnarray}} 
\newcommand{\eea}{\end{eqnarray}}

\newcommand{\bmp}{\noindent\begin{minipage}{16cm}}

\definecolor{emph}{rgb}{0.18,0.18,0.60}
\definecolor{emph1}{rgb}{0.6,0.067,0.013}
\definecolor{emph2}{rgb}{0.18,0.18,0.60}

\begin{document}

\title{ Accurate simulation of the finite density lattice Thirring model }

\author{%
\firstname{Jarno} \lastname{Rantaharju} \inst{1} \thanks{\email{jmr108@phy.duke.edu}} 
}

\institute{%
Duke University,
}

\abstract{
We present a study of the finite density lattice Thirring model in 1+1 dimensions using the world-line/fermion-bag algorithm. The model has features similar to QCD and provides a test case for exploring the accuracy of various methods of solving sign problems. In the massless limit and with open boundary conditions we show that the sign problem is an artifact of the auxiliary field approach and is completely eliminated in the fermion bag approach. With periodic boundary conditions the sign problem is mild in the fermion bag method. We present accurate results for various quantities in the model that can be used as a benchmark for comparison with other methods of solving sign problems.
}

\maketitle


\section{Introduction} \label{intro}

In lattice field theories with finite chemical potential, Monte Carlo methods are often hindered by the sign problem. The partition function may not be strictly positive and cannot be interpreted as a probability density. The problem is encountered in QCD, among other models.
Several general methods for solving sign problems are currently being studied, including the complex Langevin method and Monte Carlo simulations on Lefschetz thimbles.

In order to verify the accuracy of these methods, it is necessary to compare them to accurate data in models where the sign problem can be solved. Several exactly solvable one dimensional models have been used for this purpose. \cite{Alexandru:2015sua,Mukherjee:2014hsa,Fujii:2015bua,Tanizaki:2015rda, Fujii:2015vha,Fujii:2015rdd,Alexandru:2015xva} Two dimensional models offer more complex test cases that can more accurately represent the complex physics of target models, but may not be easily solvable.

The Thirring model in 1+1 dimensions provides a physically significant test case with features similar to QCD.
The model was studied in \cite{Alexandru:2016ejd} using a holomorphic flow to produce an integration contour similar to Lefschetz thimbles.
It is asymptotically free and has a well defined continuum limit at the Gaussian fixed point. It includes a massless boson and a fermion with a dynamically generated mass, reminiscent of the pions and baryons in  QCD.
However, these features do not result from a spontaneous symmetry breaking as in QCD. Spontaneus breaking of continuous symmetries is forbidden in two dimensional systems. Instead the model is critical at the massless limit \cite{Witten:1978qu}.

We use the fermion bag algorithm \cite{Chandrasekharan:2009wc} to simulate the massless 1+1 dimensional Thirring model and provide benchmark data on the fermion number at finite density.
We show that the sign problem is absent with open spatial boundary conditions using a world line representation. The world line representation also provides an efficient worm algorithm for producing fermion bag configurations. Using open boundary conditions, we provide data with large lattice sizes and study critical scaling in the model.

\section{Numerical Studies} \label{model}

The massless staggered Thirring model is defined by the action
\begin{align}
S &= \sum_{x,y} \bar \chi_x  D^{KS}_{x,y}(\mu) \chi_y - U \sum_{x,\nu} \bar\chi_x \chi_x \bar\chi_{x+\nu} \chi_{x+\nu}\\
 D^{KS}_{x,y}(\mu) &= \frac 12 \eta_{x,\nu} e^{\mu \delta_{\nu,0} } \delta_{y,x+\nu}  - \frac 12 \eta_{x,\nu}^\dagger e^{-\mu \delta_{\nu,0} } \delta_{y,x-\nu},
\end{align}
where $\mu$ is the chemical potential. 
The action can be expressed equivalently using an auxiliary gauge field $A_{x,\mu}$
\begin{align}
S &= \sum_{x,y} \bar \chi_x D_{x,y}(\mu) \chi_y + \sum_{x,\nu} \frac{N_F}{g^2}\left ( 1-\cos A_{x,\nu} \right ) \\
D_{x,y}(\mu)  &=  \sum_{\nu} \frac {\eta_\nu} {2} \left (  e^{i A_{x,\nu} + \mu \delta_{\nu,0} } \delta_{x+\nu,y} - e^{-i A_{y,\nu} - \mu \delta_{\nu,0} } \delta_{x-\nu,y} \right )
\end{align}
This form of the action was used in \cite{Alexandru:2016ejd}. The coupling $g$ is related to $U$ as
\begin{align}
U= \frac 14 \left( \frac{ I_{0}\left(\frac{N_F}{g^2}\right) }{ I_{1}\left(\frac{N_F}{g^2}\right ) } \right)^2-\frac 14,
\end{align}
where $I_0$ and $I_1$ are modified Bessel functions.

\subsection{Fermion Bag Representation} \label{reps}

\begin{figure} \center
\includegraphics[width=0.45\linewidth]{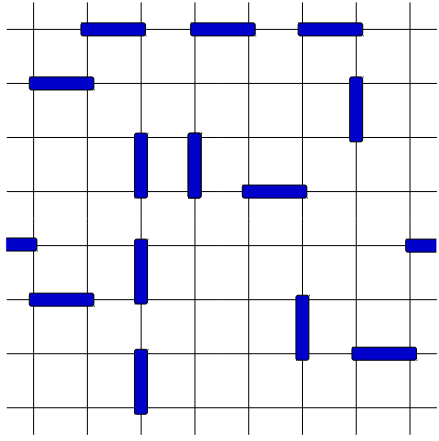} 
\caption{ A diagrammatic representation of a possible fermion bag configuration. Occupied dimers $d_{x,\nu}=1$ are represented as blue bars covering sites $x$ and $x+\hat\nu$. The configuration consists of two fermion bags, isolated from each other by the dimers.  }
\label{linkconfig}
\end{figure}

The weight of a configuration of fermion fields is given by the partition function
\begin{align}
Z = \int d\bar\chi d\chi e^{-S}
\end{align}
The fermion bag \cite{Chandrasekharan:2009wc} representation is found be expanding the partition function
\begin{align}
Z &= \int d\bar\chi d\chi e^{-\sum_{x,y} \bar \chi_x  D^{KS}_{x,y}(\mu) \chi_y} \prod_{x,\nu} \left ( 1+ U  \bar\chi_{x} \chi_{x} \bar\chi_{x+\nu}\chi_{x+\nu} \right )\\
&= \sum_{[d]} U^{N_d} \int d\bar\chi d\chi \left ( \bar\chi_{x} \chi_{x} \bar\chi_{x+\nu}\chi_{x+\nu} \right )^{d_{x,\nu}}  e^{-\sum_{x,y\in [f]} \bar \chi_x  D^{KS}_{x,y} \chi_y},
\end{align}
where $d_{x,\nu}$ is a new dimer variable that takes values $0$ and $1$, [d] is the set of possible dimer configurations and $[f]$ is the set of sites not occupied by a dimer. An example configuration is shown in Figure~\ref{linkconfig}. Each dimer occupies two sites, $x$ and $x+\hat\nu$.

It is now straightforward to integrate over the fermion fields on occupied sites
\begin{align}
 Z &= \sum_{[d]} U^{N_d} \int_{[f]} d\bar\chi d\chi e^{-\sum_{x,y\in [f]} \bar \chi_x  D^{KS}_{x,y} \chi_y}\\
 &= \sum_{[d]} U^{N_d} \det\left( W([f],\mu) \right ),
\end{align}
where the matrix $W([f],\mu)$ is constructed out of the fermion matrix $D^{KS}(\mu)$ by projecting to the space of unoccupied sites $[f]$. The weight of a given configuration of dimers is then
\begin{align}
 Z([d]) = U^{N_d} \det\left( W([f],\mu) \right ).
\end{align}

Configurations of $d_{x,\nu}$ can be generated using heatbath updates.  There are two possible updates, adding and removing a dimer. The determinant of the suggested configuration is calculated and the new configuration is accepted with the probability $\min(\left|Z([d'])/Z([d])\right|,1)$, where $Z([d'])$ is the partition function of the new configuration. If negative weights are encountered, the observables must be reweighed to the actual probability density
\begin{align}
\ev{O} &= \frac { \sum_{[d]} \textrm{sign}\left (Z([d]) \right ) ~ O  }{ \sum_{[d]} \textrm{sign}\left (Z([d]) \right )  }.
\end{align}

\begin{figure}
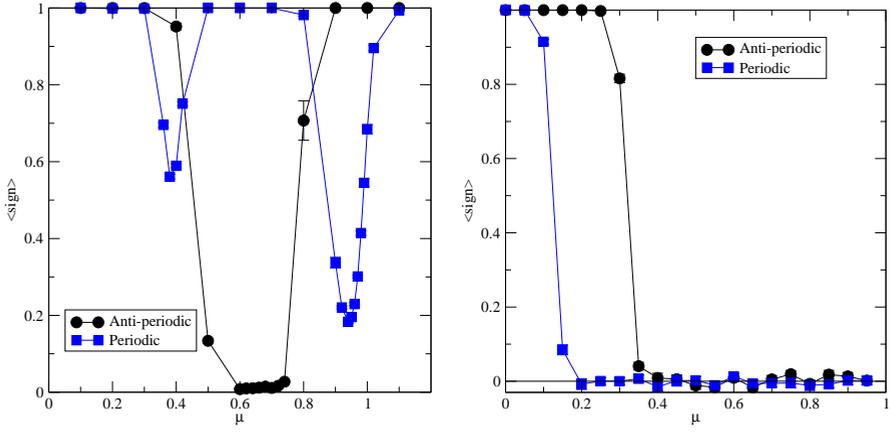
\center
    \begin{subfigure}{
        \includegraphics[width=0.4\textwidth]{sign}
        }
    \end{subfigure}
    \begin{subfigure}{
        \includegraphics[width=0.4\textwidth]{sign_gauge}
        }
    \end{subfigure}
    \caption{ The average sign with $L_X=6$ and $L_T=48$ with periodic and anti-periodic boundary conditions. Left: Using the fermion bag algorithm with $U=0.3$. Right: Using the auxiliary field representation with $g=1.05$.}
    \label{signs}
\end{figure} 

\begin{figure}
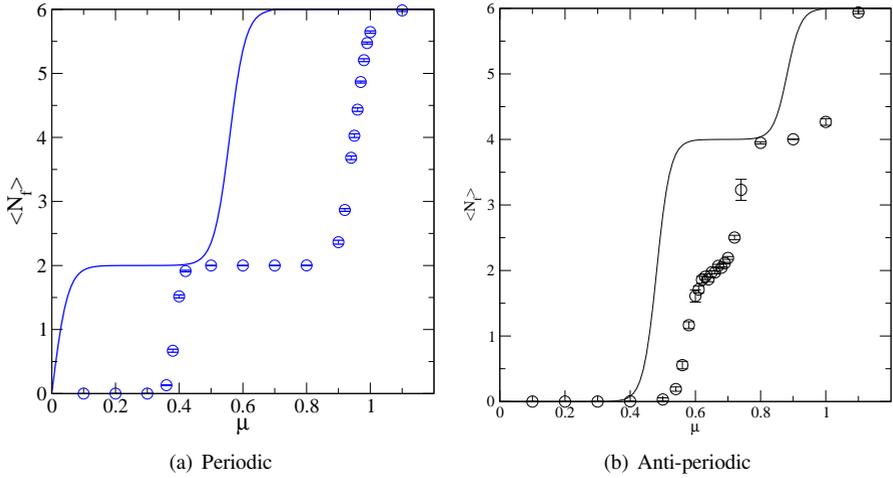
\center
    \begin{subfigure}[Periodic]{
        \includegraphics[width=0.4\textwidth]{sym_charge}
        }
    \end{subfigure}
    \begin{subfigure}[Anti-periodic]{
        \includegraphics[width=0.4\textwidth]{asym_charge}
        }
    \end{subfigure}
    \caption{ The fermion number $\ev{N_f}$ with $L_X=6$ and $L_T=48$. The solid lines shows the values at $U=0$ and the circles at $U=0.3$.}
    \label{nf-periodic}
\end{figure} 

In order to measure the severity of the sign problem at nonzero chemical potential, we measure the average sign of the partition function. The results are shown in Figure~\ref{signs} in the left panel. The right panel shows the average sign in the auxiliary field representation for comparison. In both cases we use both periodic and anti-periodic boundary conditions in spatial directions and $L_X=6$ and $L_T=48$. In the fermion bag model we use the coupling $U=0.3$ and in the auxiliary field model we use the comparable $g=1.05$. The temporal boundary condition are always anti-periodic.

The sign problem is mild with periodic boundary conditions. The average sign differs from $1$ around two values of the chemical potential, $\mu=0.4$ and $\mu=1$. With anti-periodic boundary conditions, the sign problem is severe in a wider region between $\mu=0.4$ and $\mu=0.9$. It can still be overcome using large statistics in these small volumes.

In a finite volume system the energy levels of possible states are quantized. As the chemical potential is increased the fermion number increases in a step like fashion as the chemical potential exceeds these energy levels.
An adequate solution to the sign problem should reproduce this behavior.
In Figure~\ref{nf-periodic} we show the fermion number
\begin{align}
\ev{N_f} &= \ev{\sum_{x\in S} \frac {\eta_{x,\alpha}} 2 \left [ e^\mu \bar\psi_x \psi_{x+\alpha} - e^{-\mu}\bar\psi_{x+\alpha} \psi \right ] }
\end{align}
As a function of the chemical potential.
The step-like behavior is reproduced in both cases and we find a shift in the energy of the lowest state due to the interaction term, consistent with a dynamically generated mass. In the anti-periodic case we find a region of non-trivial behavior corresponding to the region of small average sign. We also notice that the regions with average sign different from $1$ correspond to the jumps in the fermion number in both cases.

\subsection{World Line Representation} \label{sign}

The sign problem is notably absent when the spatial boundary conditions are open,
\begin{align}
&\bar \chi_{x=-1} = \chi_{x=-1} = \bar \chi_{x=N_x} = \chi_{x=N_x} = 0,
\end{align}
and the temporal boundary conditions are anti-periodic.
This result can be shown exactly using a world line representation of the partition function. The method also allows a fast worm algorithm for producing fermion bag configurations.

The world line representation is an expansion of the exponential form of the fermion determinant
\begin{align}
\det\left( W([f],\mu) \right )
&= \prod_{x\in[f]} \left ( \int d\bar\chi_x d\chi_x \right )
\times e^{-\sum_{x \in [f]} \left ( \frac 12 \eta_{x,\nu} e^{\mu \delta_{\nu,0} }  \bar \chi_x \chi_{x+\nu}  - \frac 12 \eta_{x,\nu}^\dagger e^{-\mu \delta_{\nu,0} }  \bar \chi_{x+\nu} \chi_{x} \right ) }\\
&=  \prod_{x\in[f]} \left ( \int d\bar\chi_x d\chi_x \right ) \prod_{x,x+\nu \in [f]}
 \left ( 1 - \frac 12 \eta_{x,\nu} e^{\mu \delta_{\nu,0} }  \bar \chi_x \chi_{x+\nu} + \frac 12 \eta_{x,\nu}^\dagger e^{-\mu \delta_{\nu,0} }  \bar \chi_{x+\nu} \chi_{x} \right ) \\
&= \sum_{[l]}  \prod_{loop \in [l]} \left(-\prod_{l_{x,\alpha} \in loop}   e^{l_{x,\alpha}\mu \delta_{\alpha,0}} \frac {l_{x,\alpha} \eta_{x,\alpha}}2 \right ),
\end{align}
where $l_{x,\alpha}=0,\pm1$ are new directed link variables, with $l_{x,\alpha}=+1$ representing a term with $\bar \chi_x \chi_{x+\nu}$ and $l_{x,\alpha}=-1$ a term with $
\bar\chi_{x+\nu} \chi_{x}$. 
Since the Grassmann integral is only nonzero if we have exactly one instance of $\bar\chi_x$ and $\chi_x$ at all sites, the links must form closed loops and cannot overlap.
In the last step we performed the Grassmann integral over all fermion fields and arranged the sum over the link variables into loops.

\begin{figure} \center
    \begin{subfigure}{
        \includegraphics[height=3cm]{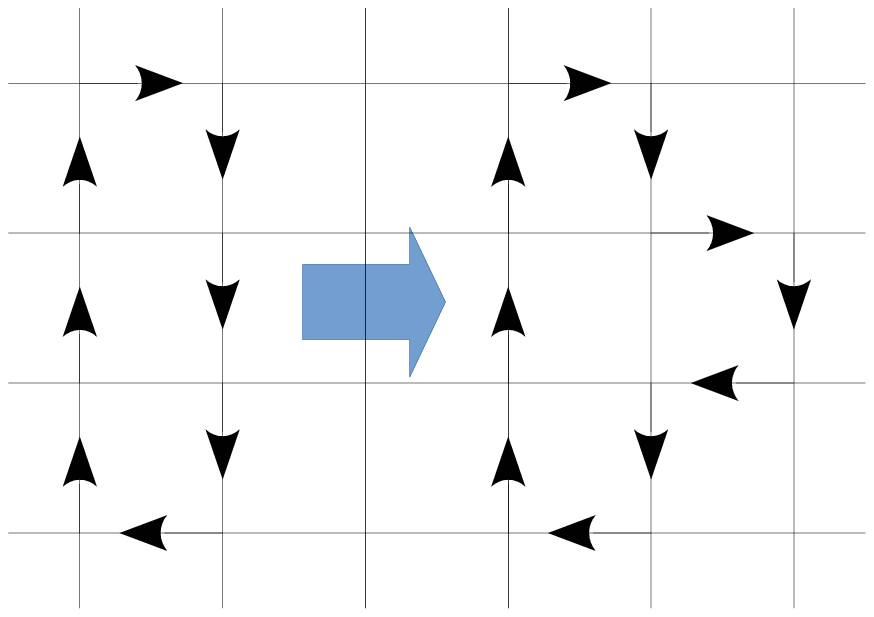}
    }
    \end{subfigure}
    \begin{subfigure}{
        \includegraphics[height=3cm]{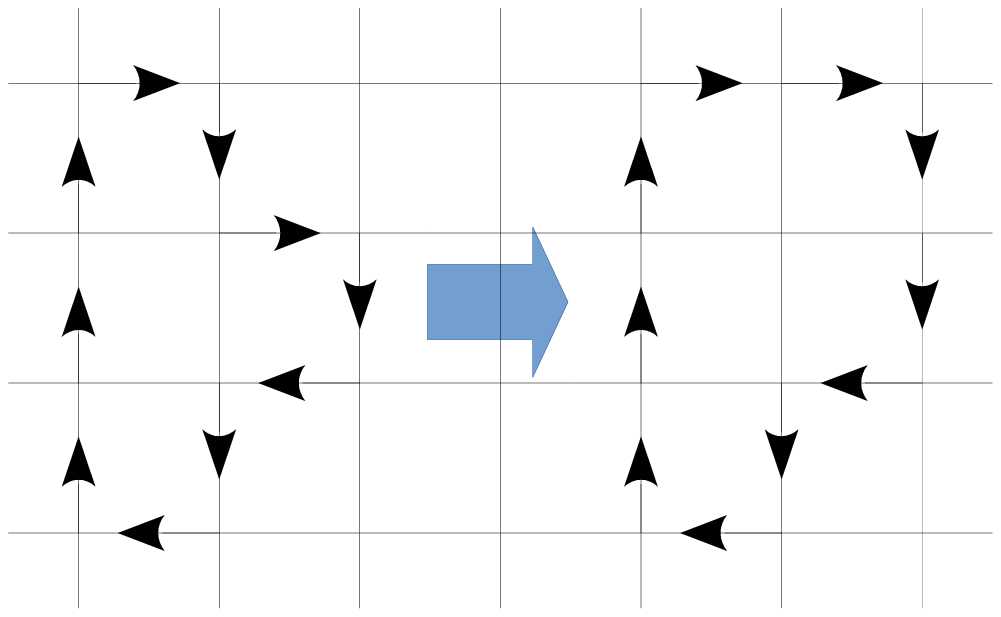}
    }
    \end{subfigure}
    \caption{Two deformations that allow the enumeration of all loops. Deformation (a) on the left on the left does not change the sign of the enclosed volume. Deformation (b) on the right on the right changes both the sign and the enclosed volume. }
    \label{deformations}
\end{figure}

The weight of a configuration is then a product of the weights of the loops in the configuration and the weights of the dimers. We will show that all allowed loops have positive weight given open boundary conditions. Consider first loops that do not wrap around the time direction. We can enumerate all such loops by starting from a simple loop and using the two deformations shown in Figure~\ref{deformations}.

Deformation (a) replaces a link with a staple or a staple with a link and does not change the sign of the loop. Deformation (b) replaces a corner with an inverted corner and does change the sign of the loop. It also also changes the volume enclosed by one. Repeating these two step we find that all loops that enclose an even number of sites have the same sign. Since the simple two-site loop has a positive sign, any loop that encloses an even volume also has a positive sign. Since odd volumes cannot be filled with dimers or loops, only even volumes are allowed.

The same argument applies for time wrapping loops that wrap around the time direction. These loops split the lattice into two volumes, one on each side.
Starting from a loop pointing directly up or directly down, we can enumerate all other loops. The weight of this starting loop is positive and since only even volumes are allowed, the weights of all allowed loops are positive.
The argument does not apply if a mass term in included in the action or when the spatial boundary conditions are periodic or anti-periodic. In these cases it is trivial to construct allowed loops with a negative weight.

\begin{figure} \center 
\includegraphics[height=0.2\linewidth]{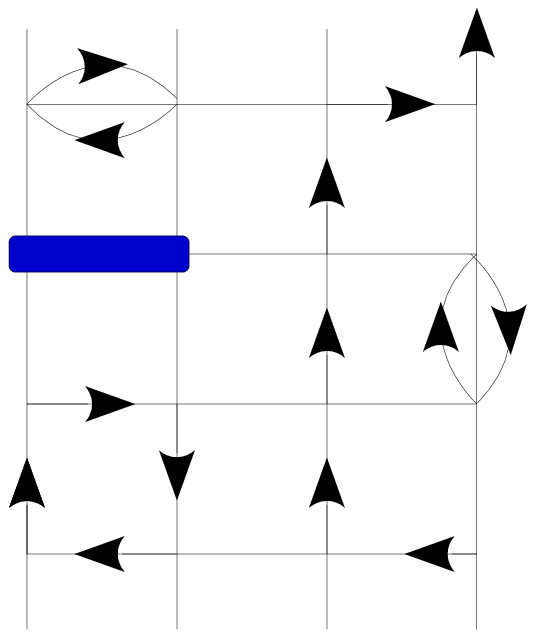}
\includegraphics[height=0.2\linewidth]{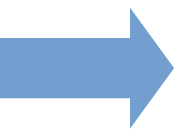}
\includegraphics[height=0.2\linewidth]{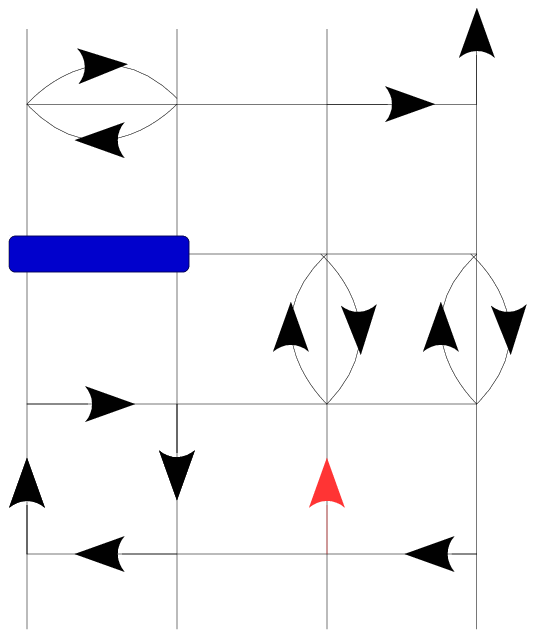}
\includegraphics[height=0.2\linewidth]{{wluarrowright}.eps}
\includegraphics[height=0.2\linewidth]{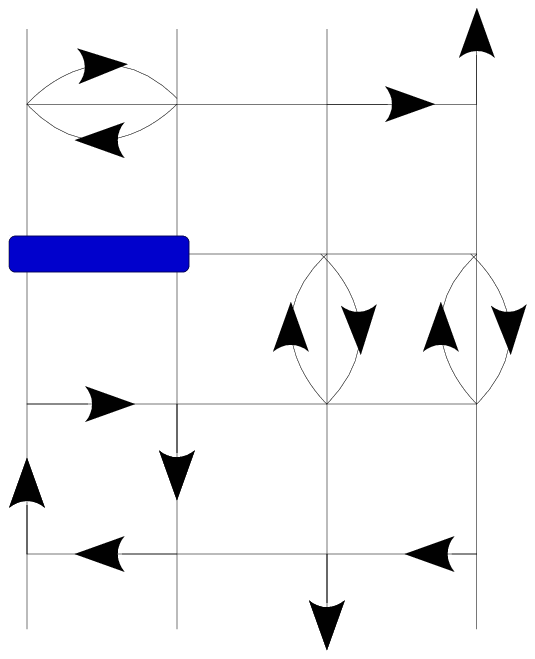}\\\vspace{0.2cm}
\includegraphics[height=0.2\linewidth]{{wlustate1}.eps}
\includegraphics[height=0.2\linewidth]{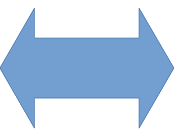}
\includegraphics[height=0.2\linewidth]{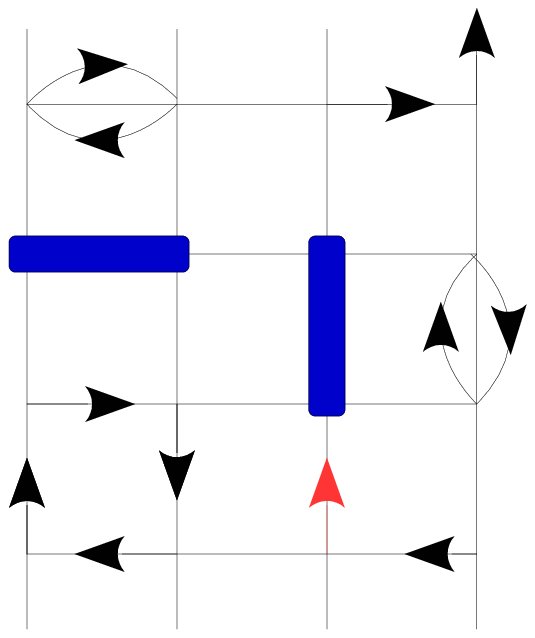}
\label{wlupdates}
\caption{An example of the worm update for generating world line and fermion bag configurations. The fermion links are shown as black arrows. The red arrow marks the extra arrow pointing at the head of the worm. In the first step, one of the arrows is pointed to a new direction, creating a defect. In the second step, the extra arrow is randomly pointed to a new direction. The worm closes when the new link points to the starting site and no defect is created. The last two configurations demonstrate adding or removing a dimer using a heatbath step.}
\label{worm_example}
\end{figure}

Since all possible configurations of loops have positive weight, it is straightforward to generate configurations within the world line representation using a worm update. The update begins by changing a random link from $l_{x,\alpha}$ to $l_{x,\alpha'}$ with the probability
\begin{align}
P=\min\left(1, e^{l_{x,\alpha'}\mu \delta_{\alpha',0} - l_{x,\alpha}\mu \delta_{\alpha,0} } \right).
\end{align}
If accepted, the update creates two defects. There is now a site with no link pointing to it and a site with two links pointing to it. The first is the tail and the second the head of the worm. The new link is kept and a new direction is suggested for the old link pointing at the head. With this update the head propagates to a new site. The process is repeated until a link pointing at the tail is accepted.

The set of dimers may be updated using a simple heatbath step. Loops consisting of two site can be turned into dimers with the probability $\min\left(1,4U\right)$ and dimers into loops with the probability $\min\left(1,1/4U\right)$. An example of the worm update is shown in Figure~\ref{worm_example}.

\begin{figure}
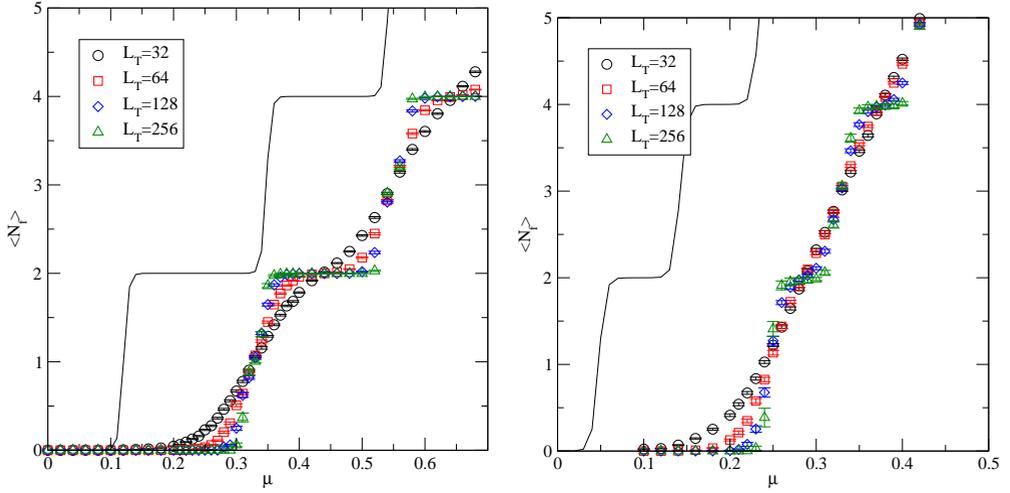

    \centering
    \begin{subfigure}{
        \includegraphics[width=0.45\textwidth]{charge_12}
    }
    \end{subfigure}
    \begin{subfigure}{
        \includegraphics[width=0.45\textwidth]{charge_32}
    }
    \end{subfigure}
    \caption{ Fermion number at $U=0.3$ and $L_X=12$ (left) and $32$ (right) }
    \label{open_nf}
\end{figure} 

It is now straightforward to produce data even on large lattices. As before, we reproduce the step-like behavior of the fermion number on finite lattices. 
The fermion number in the world line formalism is simply
\begin{align}
\ev{N_f} &= \ev{ \sum_{x\in S} l_{x,\hat t} - l_{x+\hat t,-\hat t} }
\end{align}
Figure \ref{open_nf} shows the fermion number at $U=0.3$ with several lattice sizes up to $L_X=32$ and $L_T=256$. The step-like behavior of the fermion number is clearly visible.
At nonzero coupling $U$, the steps are shifted to a larger $\mu$, providing evidence of the dynamically generated fermion mass.

\begin{figure}
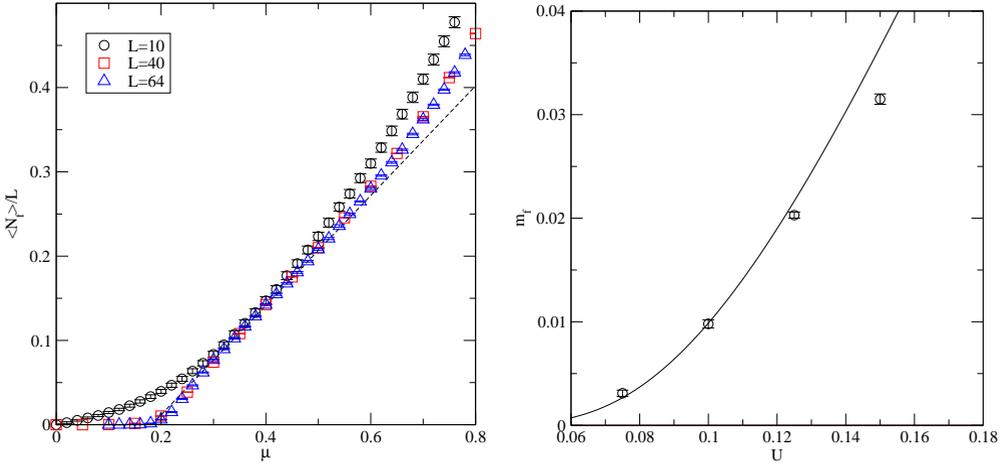

    \begin{subfigure}{
        \includegraphics[width=0.45\textwidth]{charge_square}
    }
    \end{subfigure}
    \begin{subfigure}{
        \includegraphics[width=0.45\textwidth]{mf_beta_scaling}
    }
    \end{subfigure}
    \caption{Left: The fermion number density of square lattices. The dashed line shows a fit to the linear prediction at infinite volume. Right: A comparison of the fermion mass and the expected asymptotic scaling. }
    \label{square}
\end{figure}

The model is asymptotically free and at small $U$, in the continuum limit, it should be possible to connect with perturbation theory. This requires large lattice sizes to reduce finite volume effects. 
At infinite volume but small coupling $U$, the fermion mass scales as
\begin{align}
m_f = c e^{\frac{-2\pi}{b_0U}},
\end{align}
where $b_0$ is the one-loop coefficient of the $\beta$-function.

We measure the fermion mass using large square lattices at small $U$. In this case the fermion number density $\ev{n}=\ev{N_f}/L$ increases linearly with the chemical potential
\begin{align}
\ev{n} = \max\left(0, c\left( \mu-m_f \right) \right ).
\end{align}
We show measurements of the fermion number density at $U=0.3$ in Figure~\ref{square} on the left. The fermion mass is $m_f \approx 0.2$ in this case. On the right in Figure~\ref{square} we show the fermion mass at small coupling and large lattice size, up to $L=1024$, together with the asymptotic scaling with $b_0=16$ and $c=0.49$. We find qualitative agreement with the perturbative expectation.

\section{Conclusions} \label{conclusions}

We study the massless 1+1 dimensional lattice Thirring model with finite chemical potential using the fermion bag formulation. With periodic and anti-periodic boundary conditions we find a milder sign problem than in the auxiliary field formalism. We are able extract accurate measurements with small lattice sizes.

We show that the sign problem is removed with open boundary conditions in the spatial direction using a world line formalism. The world line expansion naturally leads to an efficient worm update for generating fermion bag configurations and allows accurate measurements with very large lattice sizes.
We then study the model at multiple lattice sizes, robustly reproducing the step-like behavior of the fermion number and showing that a fermion mass is generated dynamically by the interaction. At small coupling and large lattice size we find a qualitative match to the expected asymptotic behavior of the fermion mass.

\section{Acknowledgements} \label{acknowledgements}
This study is done in collaboration with S. Chandrasekharan and V. Ayyar.
We thank A. Alexandru and P.F. Bedaque for useful discussion. This work was supported by the U.S. Department of Energy, Office of Science, Nuclear Physics program under Award Number DE-FG02-05ER41368.

\bibliography{lattice2017}

\end{document}